\documentclass[letterpaper]{jpconf}
\usepackage{graphicx}
\begin{document}
\title{The Universal Edge Physics in Fractional Quantum Hall Liquids}

\author{Zi-xiang Hu$^1$, R. N. Bhatt$^1$, Xin Wan$^2$, Kun Yang$^3$}
\address{$^1$Department of Electrical Engineering, Princeton University, Princeton, New Jersey
08544, USA}
\address{$^2$ Zhejiang Institute of Modern Physics, Zhejiang University, Hangzhou 310027, P.R. China}
\address{$^3$National High Magnetic Field Laboratory and Department of Physics, Florida State University, Tallahassee, Florida 32306, USA}
\ead{zihu@princeton.edu}

\begin{abstract}
The chiral Luttinger liquid theory for fractional quantum Hall edge transport predicts universal power-law behavior 
in the current-voltage ($I$-$V$) characteristics for electrons tunneling into the edge. However, it has not been
unambiguously observed in experiments in two-dimensional electron gases based on GaAs/GaAlAs heterostructures or quantum wells. One plausible cause is the fractional quantum Hall edge reconstruction, which introduces non-chiral edge modes. The coupling between counterpropagating edge modes can modify the exponent of the $I$-$V$ characteristics. By comparing the $\nu=1/3$ fractional quantum Hall states in modulation-doped semiconductor devices and in graphene devices, we show that the graphene-based systems have an experimental accessible parameter region to avoid the edge reconstruction, which is suitable for the exploration of the universal edge tunneling exponent predicted by the chiral Luttinger liquid theory. 
\end{abstract}

\section{Introduction}

The fractional quantum Hall (FQH) effect has been observed in GaAs-based semiconductor heterostructures and quantum wells~\cite{Tsui} for almost 30 years. 
The fractional quantum Hall states have a lasting fascination due to their nontrivial topological properties, which lie beyond the paradigm of the Fermi-liquid theory, 
and the potential applications in topological quantum computation~\cite{tqc, tqc1, tqc2}. 
Probing excitations at the edge of a FQH droplet is the most accessible approach for the detection of the topological properties in the bulk thanks 
to the edge-bulk correspondence in the topological system. The edge excitations are gapless in the thermodynamic limit, contrast to gapped 
quasihole/quasiparticle excitations in the bulk. Owing to the existence of edge states, current exists between two contacts connected by an edge 
channel as electrons can be injected into or removed from the FQH edge without costing energy. The standard theory for the FQH edge physics is the chiral Luttinger liquid 
(CLL) theory~\cite{Wen}. The theory predicts that a FQH droplet exhibits power-law bahavior in the $I$-$V$ characteristics ($I \propto V^{\alpha}$)
 when electrons are tunneling through a barrier into the edge, e.g., from a three-dimensional Fermi liquid. For the celebrated Laughlin state at a 
filling factor $\nu = 1/3$ in the lowest Landau level, the CLL theory predicts a tunneling exponent $\alpha = 1 / \nu = 3$~\cite{Wen}. However, the 
exponent $\alpha$ measured in experiments are sample dependent with a numerical value smaller than 3, which are nonuniversal~\cite{grayson, amchang,amchang01, chamon}. 
One of the possible causes of this discrepancy is existence of counterpropagating edge modes, which result from edge reconstruction~\cite{chamon,xinprl,yangprl03}. The 
edge reconstruction in the electron density profile manifests the competition between electron-electron interaction, which tends to spread out the electron density, 
and the electrostatic confinement at the edge, which holds the electrons in the interior of the sample. Numerical calculations~\cite{xinprb,jain} indicate that the 
edge reconstruction in FQH states is unavoidable in the state-of-art GaAs-based samples based on modulation doping, which greatly enhances sample mobility but 
unintentionally introduces a large region of fringe field at the edge that favors edge electron redistribution. This effect hinders the observation of the universal behavior in edge transport.

In recent years new two-dimensional materials, such as graphene, emerged. The FQH effect has already been observed in graphene~\cite{Kirill,XDu,dean}. However, integer and fractional quantum Hall effects in graphene differ from those in normal semiconductor-based devices, because electrons in graphene obey the relativistic Dirac equation, instead of the nonrelativistic Schr\"odinger equation often encountered in solid state materials. Here we compare the FQH states in both GaAs-based samples and in graphene samples with the emphasis on the difference of the competition at the edge. We find that there is an experimental accessible parameter region in graphene-based samples, in which edge reconstruction can be avoided and, therefore, universal FQH edge physics can be observed. The universality is reflected in the calculation of the equal-time edge Green's function, which is related to the $I$-$V$ characteristics in edge tunneling.

\begin{figure}
 \includegraphics[width=9cm]{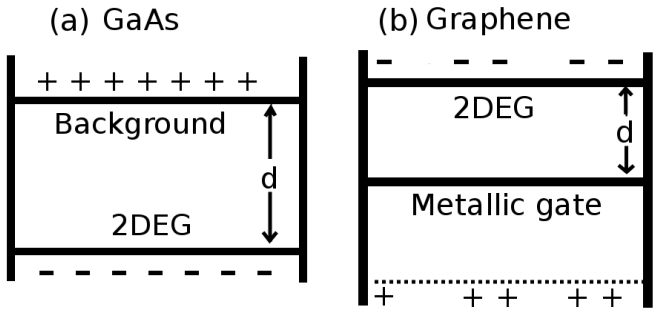} 
\begin{minipage}[b]{6.6cm}\caption{\label{setup} Simplified models to mimic the experimental setups in (a) GaAs-based and (b) graphene-based systems. The GaAs model consists of an electron layer and a homogeneous impurity charge layer, which confines electrons. The graphene model consists of layered Dirac electrons with an open edge and their image charge due to metallic gate; both confines electrons.}
\end{minipage}
\end{figure}
 
\section{FQH states in GaAs-based systems}
An experimental sample based on a modulation-doped GaAs/GaAlAs heterostructure~\cite{pfeiffer} can be simplified to a 
two-dimensional electron layer, which is located at the interface between GaAs and GaAlAs, and a homogeneous positive background charge layer, which originates from the remote dopants. This is illustrated in Fig.~\ref{setup} (a). The confinement potential for electrons due to background charge competes with the Coulomb interaction between electrons. This competition is the driving force for edge reconstruction.  The density of the background charge $\sigma$ is determined by the electron filling factor $\nu$ due to charge neutrality and, therefore, the background confinement potential is a single-body potential in the FQH Hamiltonian. The complete Hamiltonian can be written as
\begin{equation}
\label{Hamiltonian}
 H = \frac{1}{2}\sum_{mnl}V_{mn}^l c_{m+l}^+c_n^+c_{n+l}c_m +
     \sum_m U_m c_m^+ c_m,
\end{equation}
where the $c_m^+$ is the electron creation operator for the lowest LL single electron state with an angular momentum $m$. 
\begin{eqnarray}
 V_{mn}^l = \int d^2 r_1 \int d^2r_2 \phi_{m+l}^*(r_1)\phi_n^*(r_2)\frac{e^2}{r}\phi_{n+l}(r_2)\phi_m(r_1),
\end{eqnarray}
represents the electron-electron interaction and 
\begin{equation}
U_m = e\sigma \int\int_{r_2 < R} d^2r_1 d^2 r_2\frac{1}{\sqrt{d^2 + |\vec{r_1}-\vec{r_2}|^2}}|\phi_m(r_1)|^2
\end{equation}
the confinement potential. The distance $d$ between the electron layer and the background charge layer is the parameter that controls the relative strength of the confinement potential to the electron-electron interaction. When $d$ is small, the confinement is 
strong and electrons tend to stay in the region defined by photo lithography process ($r < R$ for the disk sample we consider here). However, the confinement becomes weak when $d$ is large and thus the electrons near the edge tend to spread out, leading to edge reconstruction.

\begin{figure}
 \includegraphics[width=5cm,height=4cm]{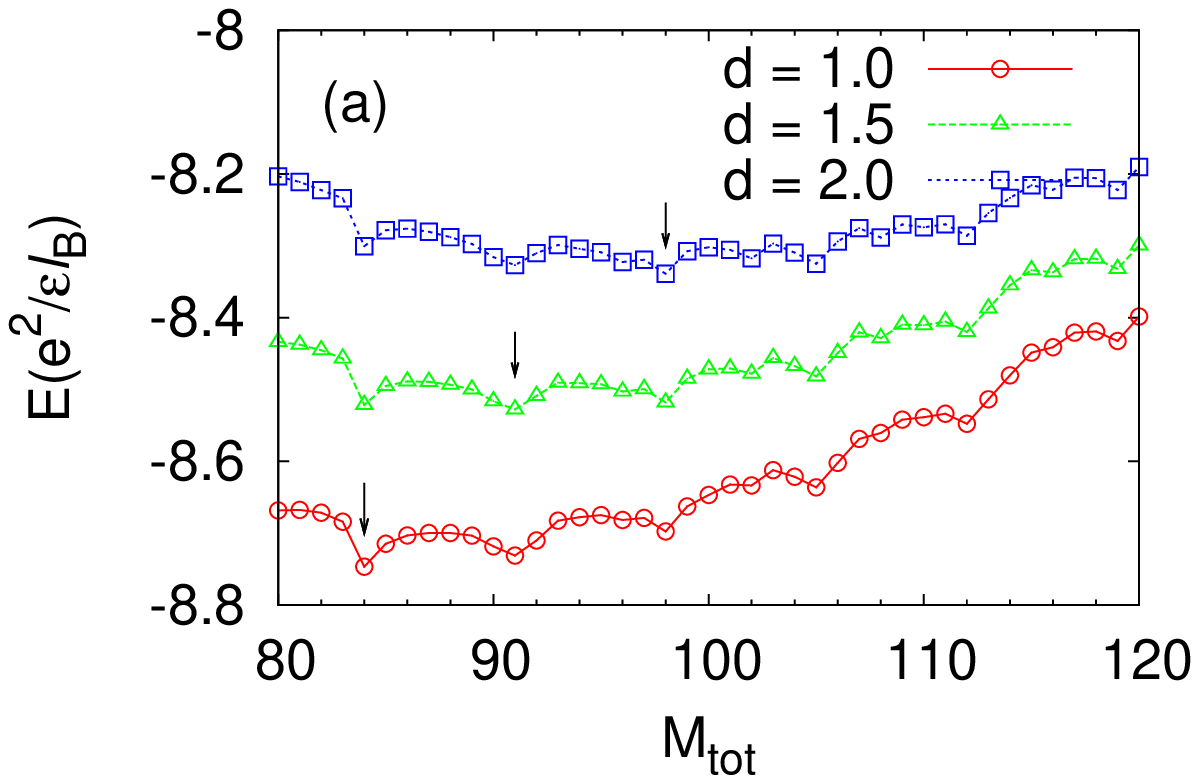}
 \includegraphics[width=5cm,height=4cm]{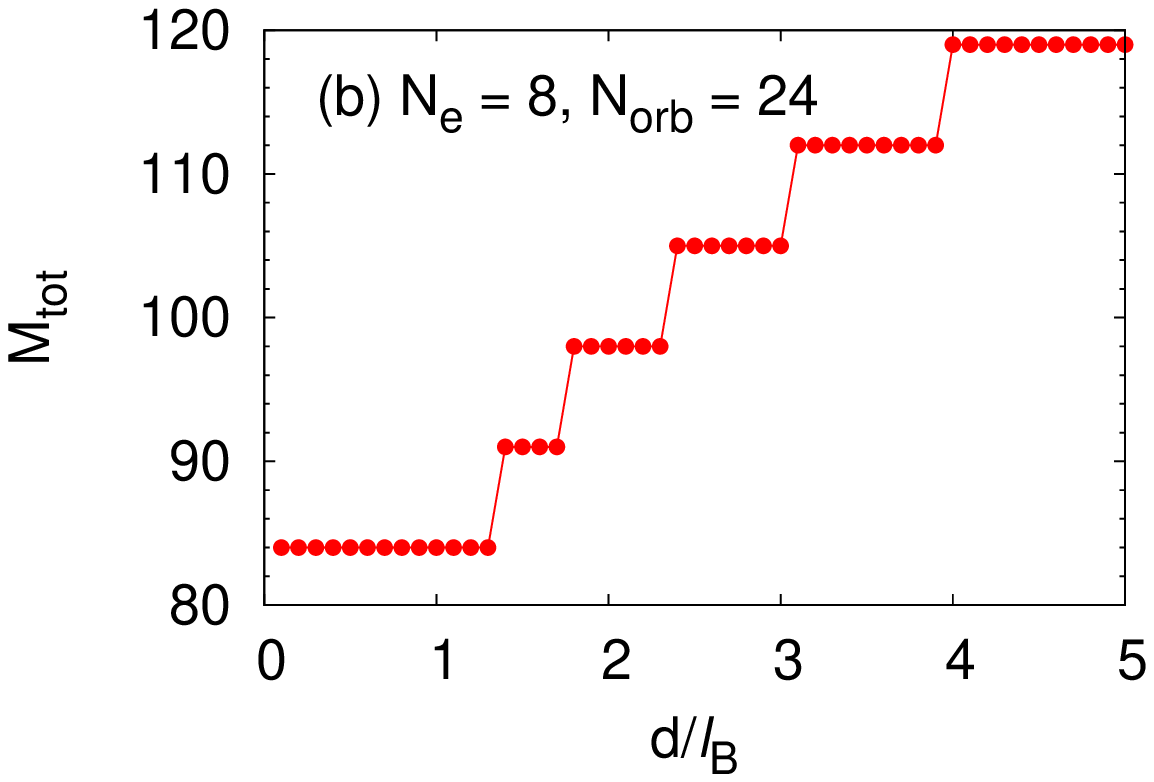} 
\begin{minipage}[b]{5.5cm}\caption{\label{gsN8} (a) The lowest energy in each total angular momentum subspace for 8 electrons in 24 orbitals with different $d$s in the GaAs model. The global ground state is labeled by an arrow. (b) The corresponding angular momentum of the global ground state as a function of $d$.}
 \end{minipage}
\end{figure}

\begin{figure}
\includegraphics[width=16cm,height=4cm]{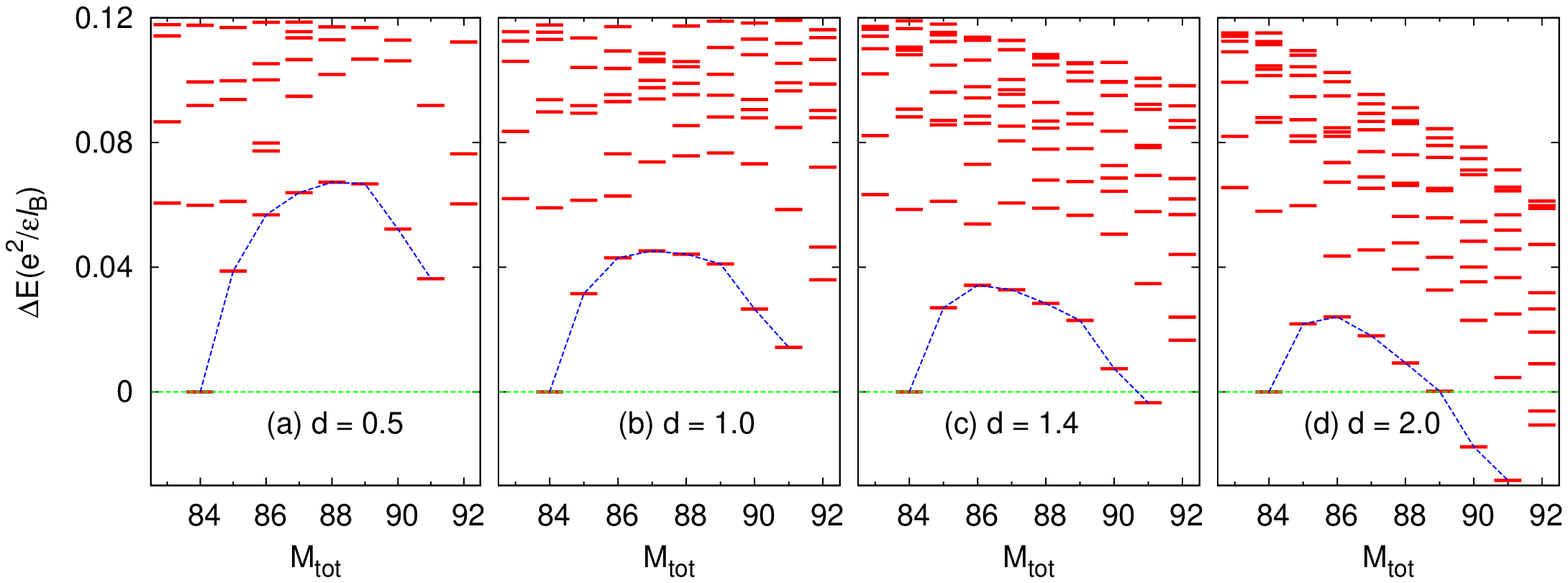}
\caption{\label{gaspec} The low-energy spectrum for 8 electrons in 26 orbitals with (a) $d=0.5\ l_B$, (b) $d=1.0\ l_B$, (c) $d=1.4\ l_B$, and (d) $d=2.0\ l_B$ in the GaAs model. Edge reconstruction, signaled by the ground state angular momentum change, occurs around $d_c \sim 1.4 \ l_B$.} 
\end{figure}

\begin{figure}
\includegraphics[width=16cm,height=4cm]{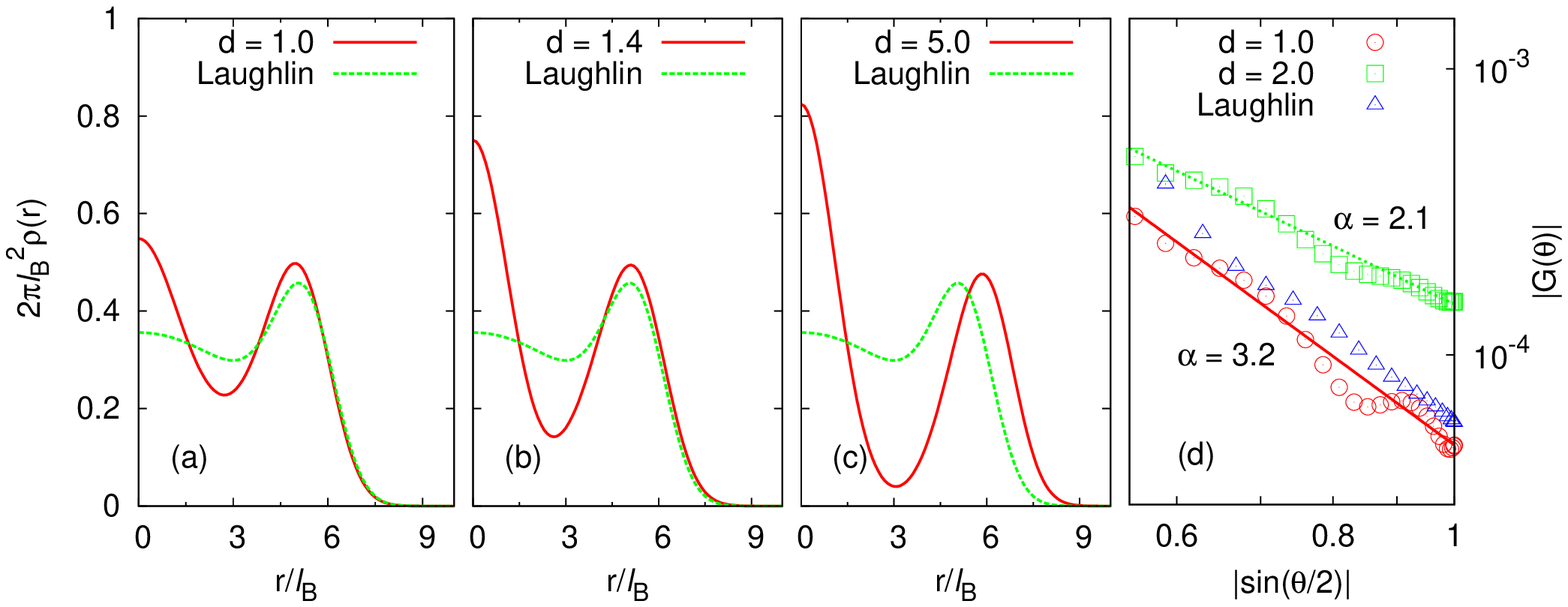}
\caption{\label{Gadensity} The electron density profile for the ground state at (a) $d=1.0\ l_B$, (b) $d=1.4\ l_B$ and (c) $d=5.0\ l_B$ in the GaAs model.  The dashed line is the density profile for the Laughlin state.
(d) The edge Green's function before and after edge reconstruction; the slopes in the log-log plot yield $\alpha = 3.2$ and 2.1, respectively.} 
\end{figure}

Because of the rotational symmetry in the geometry and the gauge we choose, the total angular momentum $M_{tot}$ is a good quantum number and thus we can diagonalize the Hamiltonian in each angular momentum subspace.
In Fig.~\ref{gsN8}(a), we plot the lowest energy in each of the total angular momentum subspace with different $d$s for 8 electrons in 24 orbitals. As discussed in Ref.~\cite{xinprl,xinprb}, when $d$ is smaller
than the critical value $d_c = 1.5 \pm 0.1\ l_B$, the global ground state has the same total angular momentum $M_{tot} = M_0 = 3N(N-1)/2$ as that of the Laughlin state. When $d > d_c$, edge electrons spread out under the weak confinement and, therefore, the total angular momentum of the global ground state increases as shown in Fig.~\ref{gsN8}(b).  To further reveal how the edge reconstruction happens with increasing $d$, we plot the low-lying energy spectrum in Fig.~\ref{gaspec} for 8 electrons in 26 orbitals with different $d$s. The larger number (26 rather than 24 or 22) of orbitals is chosen to avoid unnecessary mixing of the bulk and edge modes in a finite system. The chiral bosonic edge mode (dashed line), which describes the edge electron density deformation, is linear for small $\Delta M = M_{tot} - M_0$ but develop downward curvature as $\Delta M$ increases~\cite{xinprb}. As $d$ increases, the bosonic mode bends down further and eventually the total ground state momentum jumps from $M_{tot} = 84$ to $M_{tot} = 91$ above the critical value $d_c$; hence edge reconstruction occurs.
Fig.~\ref{Gadensity}(a)-(c) shows the evolution of the electron density profile for the ground state as $d$ increases. 
For large enough $d$, the electron density profile develops an almost detached edge piece, as shown in Fig.~\ref{Gadensity}(c), suggesting the reconstructed edge can accommodate three counterpropagating edge modes. The edge transport thus depends on the coupling of the edge modes, hence no universality can be expected except in the strong coupling limit. 

To explore the universality of the FQH edge states, we calculate the equal-time edge Green's function,
\begin{equation}
 G(\vec{r}-\vec{r'}) = \frac{\langle \psi|\Psi^+(\vec{r}) \Psi(\vec{r'})|\psi\rangle}{\langle\psi|\psi\rangle},
\end{equation}
where $\Psi^+(\vec{r})$ [or $\Psi(\vec{r})$] creates (annihilates) one electron at $\vec{r}$ at the edge of the FQH droplet. In the disk geometry with $|r| = |r'| = R$, the distance $|\vec{r} - \vec{r'}| = 2R\sin(\theta/2)$ where $\theta$ is the angle between $\vec{r}$ and $\vec{r'}$. The CLL theory predicts that in the large distance limit, the edge Green's function has an asymptotic behavior  
\begin{equation}\label{greenfunction}
 |G(\vec{r}-\vec{r'})| \sim  |\vec{r}-\vec{r'}|^{-\alpha} \propto |\sin(\theta/2)|^{-\alpha}.
\end{equation}
For the $\nu=1/3$ Laughlin state, the CLL theory predicts an exponent $\alpha = 3$. 
Fig.~\ref{Gadensity}(d) plots the edge Green's function for 8 electrons in 24 orbitals with Coulomb interaction
for two $d$s on the opposite sides of the edge reconstruction transition. 
For comparision, the edge Green's function for the variational Laughlin wavefunction is also presented. 
We find an exponent $\alpha = 3.2 \pm 0.2$~\cite{wan05} at $d = 1.0$ $l_B < d_c$, which is consistent with the CLL theory prediction. 
The exponent for the variational Laughlin wavefunction is the same as that of the Coulomb ground state before reconstruction. 
However, when $d = 2.0$ $l_B > d_c$, the exponent $\alpha = 2.1$ is noticeably smaller than $3$. This suggests that the long-range Coulomb interaction does not directly lead to nonuniversal edge behavior, but the consequence of its competition with the confinement potential, i.e., edge reconstruction, does. In a typical GaAs-based sample, $d$ is of order 1000 $\mathring{A}$, i.e., about 10 $l_B$~\cite{pfeiffer}. While the numerical calculation for such a large $d$ is not practical due to the system-size constraint, we speculate that multiple edge reconstructions can occur, leading to more than three counterpropagating edge modes. Therefore, such a device is always in the edge reconstructed regime, which does not guarantee the observation of a universal edge tunneling exponent.

\section{FQH states in graphene-base systems}
 \begin{figure}
\includegraphics[width=10cm]{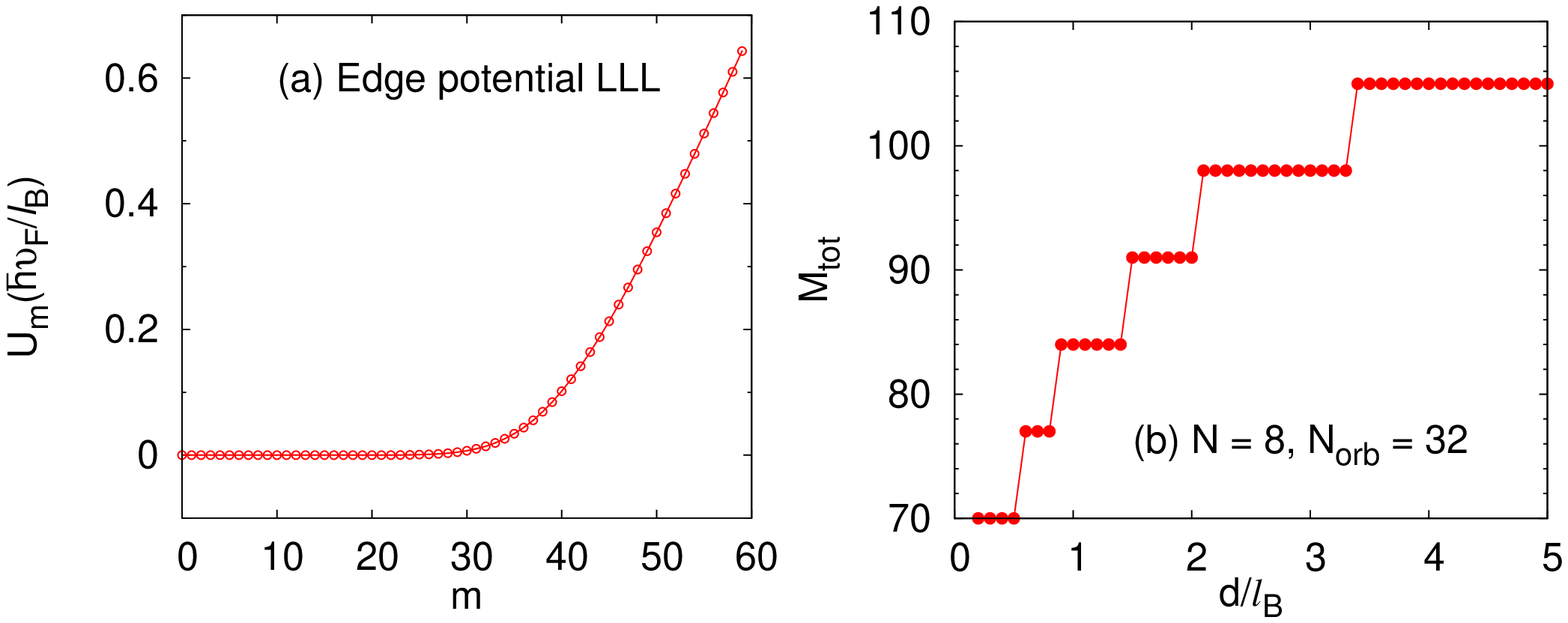}
\begin{minipage}[b]{5.5cm}
\caption{\label{graheneLL}(a) The single-particle lowest Landau level energy in the graphene model with an open edge as a function of the angular momentum in a disk sample with 60 orbitals. (b) The angular momentum of the many-body ground 
state as a function of $d$ for 8 electrons in 32 orbitals.}
\end{minipage}
 \end{figure}

The experimental setup for graphene samples is illustrated in Fig.~\ref{setup}(b) (and Ref.~\cite{zxhu}). The graphene layer is situated at a distance $d$ from a metallic gate (boron-nitride) ~\cite{dean}. Therefore, electrons have their image charge at a distance $2d$ from the graphene layer. This modifies the first term in the Hamiltonian in Eq.~(\ref{Hamiltonian}), which now contains both the electron-electron interaction and the electron-image charge interaction:
\begin{eqnarray}
 V_{mn}^l = \int d^2 r_1 \int d^2r_2 \phi_{m+l}^*(r_1)\phi_n^*(r_2)(\frac{e^2}{r} - \frac{e^2}{\sqrt{r^2+(2d)^2}})\phi_{n+l}(r_2)\phi_m(r_1).
\end{eqnarray}
The combined interaction behaves like dipole-dipole interaction with $1/r^3$ dependence at short distances and thus the interaction pseudopotentials ($V_m$s)~\cite{haldane} are dominated by those with small $m$s. 
The single-body term in Eq.~(\ref{Hamiltonian}) comes from the edge confinement potential in the graphene due to the open boundary. The solution of the Dirac equation for a semi-infinite graphene with an 
open edge at $x = 0$ satisfies equation $D_\mu(-\sqrt{2}x_c) = 0$~\cite{edgepotential} where 
$D_\mu(x)$ is the parabolic cylinder function and $x_c = k l_B$, $\mu = \epsilon^2/2$; $k$ is the wave vector parallel to the edge and $\epsilon$ the energy in units of $\hbar v_F / l_B$. 
In the disk geometry with $N_{orb}$ orbitals, the edge is located at 
$r_c = \sqrt{2N_{orb}}l_B$, thus the solution satisfies $D_\mu[-\sqrt{2}(x_c-r_c/l_B)] = 0$. Solutions are allowed for $\mu = 0, 1, 2, \cdots$ in the bulk, hence the energy for the $n$th LL is $\epsilon_n =\sqrt{2n}$. 
As plotted in Fig.~\ref{graheneLL}(a), the eigenenergy of the lowest Landau level increases rapidly as approaching the open edge, which keeps electrons from spreading out. Since in the disk geometry the $m$th orbital is 
located at $\sqrt{2m}l_B$ along the radial direction, the Landau orbitals closer to the edge are denser than those closer to the center. The edge potential influences a large portion of the orbitals in the momentum space.
Therefore, in a finite-size calculation we need to include significantly more orbitals than needed for the Laughlin state to prevent the edge potential from destroying the bulk properties. 

In Fig.~\ref{graheneLL}(b) we plot the ground-state angular momentum for 8 electrons in 32 orbitals as a function of $d$. 
Note that for small enough $d$ the ground state can have an angular momentum smaller than $M_0 = 84$, in sharp contrast to the GaAs model in Fig.~\ref{gsN8}(b), in which the Laughlin phase remains stable down to $d = 0$. This is because that the interaction is too weak to compete with the edge potential when $d$ is sufficiently small, thus electrons are repelled from the edge and the ground-state angular momentum decreases. Strictly speaking, a fair comparison requires us to use a sharp cutoff in real space in the GaAs model as well, such that the single-particle energy bends up and one needs to introduce Landau level mixing in a more complete treatment (see Ref.~\cite{xinprb} for more detail), which is irrelevant to our main results so we skip here. 
Essential to our discussion, the Laughlin phase is nearly system size independent when $N_{orb}$ is large as analyzed in Ref.~\cite{zxhu}. As pointed out there, we believe that one has a good chance of realizing the Laughlin phase in graphene without reconstructed edge for $0.5 < d/l_B < 1.5$~\cite{zxhu}. The upper bound of the window in which the unreconstructed Laughlin phase can be realized in the graphene model agrees well with that in the GaAs model~\cite{xinprl,xinprb,jain}. However, in the graphene model, e.g., on boron nitride substrate between the graphene layer and the gate, 
the parameter $d$ can be chosen as small as a few nanometers or a small fraction of $l_B$~\cite{dean}, thus the universal exponent in tunneling experiments may be easily accessible in graphene.
\begin{figure}
 \includegraphics[width=16cm, height=4cm]{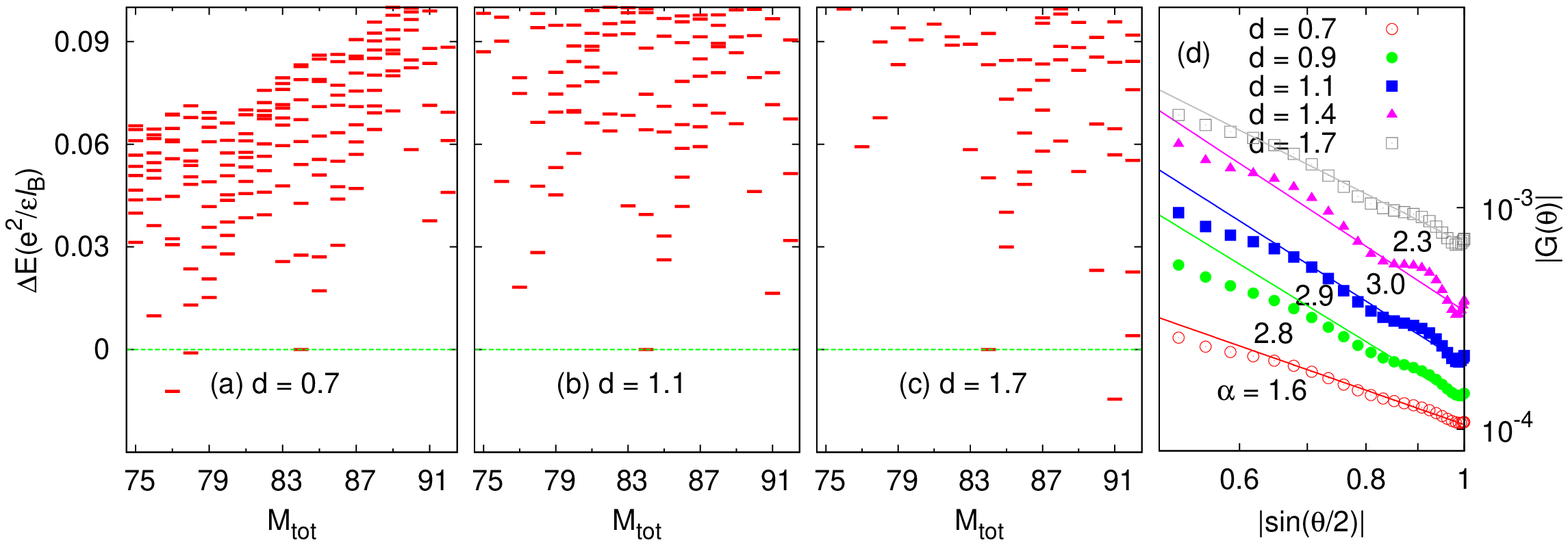}
\caption{\label{graphenegspec} (a)-(c) The low-lying energy spectrum for 8 electrons in 32 orbitals in the graphene model with various $d$s. The corresponding Laughlin state has a $M_{tot} = 84$, the same as in (b). (d) The exponent $\alpha$ in the edge Greens function of the ground states for various $d$s.}
\end{figure}

In Fig.~\ref{graphenegspec}(a)-(c) we plot the low-lying energy spectra for $d = 0.7$, 1.1, and 1.7 $l_B$, whose ground-state angular momenta are 76, 84 (Laughlin), and 91, respectively. 
In these cases the edge mode is not well separated from other states in the spectrum (similar to Fig. 8 in Ref.~\cite{xinprb}). To demonstrate the universal tunneling exponent, 
we compute the edge Green's function for 8 electrons in 32 orbitals, as shown in Fig.~\ref{graphenegspec}(d). We find that $\alpha$ for the three ground states are quite different. 
They are $1.6$ and $2.3$ for $d = 0.7\ l_B$ and $d = 1.7\ l_B$, respectively; neither is in the Laughlin phase. In these casses, the Laughlin state still suffers from the edge reconstruction 
instability, which results in charge accumulation similar to that studied earlier based on Hartree-Fock types of calculations~\cite{grapheneedge1, grapheneedge2, grapheneedge3, Castro}.
 When $d = 0.9, 1.1$ and $1.4 l_B$ within the Laughlin phase $d/l_B \in [0.9, 1.4]$ 
as shown in Fig.~\ref{graheneLL}(b), we find $\alpha \sim 2.9 \pm 0.1$. 
Therefore, the universal tunneling exponent can be expected if the Laughlin phase without reconstructed edges is confirmed in graphene.

\section{Conclusions}
We compare the FQH states at $\nu=1/3$ in models for GaAs-based and graphene-based devices. 
Although the CLL theory predicts a tunneling current $I \propto V^\alpha$ with a universal $\alpha = 3$, 
the exponent can be affected by edge reconstruction, which exists in modulation-doped GaAs-based samples. 
However, in a graphene-based device, edge reconstruction can be avoided by using a very thin dielectric between 
the graphene layer and the metallic gate. We demonstrated the universal exponent in the edge Green's function calculation. The exponent is always about 3 once the ground state is in Laughlin region and smaller than 3 outside
of the Laughlin region which consists with the CLL theory and the experiments in GaAs.

\section{Acknowledgments}
This work is supported by DOE grant No. DE-SC0002140 and the 973 Program under Project No. 2009CB929100.

\end{document}